# Heliospheric Magnetic Field 1835-2009

Leif Svalgaard[1] and Edward W. Cliver[2]

[1]Stanford University, HEPL, Cedar Hall, Via Ortega, Stanford, CA 94305-4085
[2]Space Vehicles Directorate, Air Force Research Laboratory, Hanscom AFB, MA
01731-3010

**Abstract.** We use recently acquired geomagnetic archival data to extend our long-term reconstruction of the HMF strength. The 1835-2009 HMF series is based on an updated and substantiated *IDV* series from 1872-onwards and on Bartels' extension, by proxy, of his *u*-series from 1835-1871. The new *IDV* series, termed IDV09, has excellent agreement ($R^2 = 0.98$; RMS = 0.3 nT) with the earlier IDV05 series, and also with the negative component of Love's extended (to 1905) $D_{st}$ series ($R^2 = 0.91$). Of greatest importance to the community, in an area of research that has been contentious, comparison of the extended HMF series with other recent reconstructions of solar wind *B* for the last ~100 years yields a strong consensus between series based on geomagnetic data. Differences exist from ~1900-1910 but they are far smaller than the previous disagreement for this key interval of low solar wind *B* values which closely resembles current solar activity. Equally encouraging, a discrepancy with an HMF reconstruction based on [10]Be data for the first half of the 20th century has largely been removed by a revised [10]Be-based reconstruction published after we submitted this paper, although a remaining discrepancy for the years ~1885-1905 will need to be resolved.

## 1. Introduction

In *Svalgaard and Cliver* [2005] we introduced the InterDiurnal Variability (*IDV*) index for a given geomagnetic observatory ('station') as the average difference without regard to sign, from one day to the next, between hourly mean values of the Horizontal Component, *H*, and measured one hour after midnight. The average should be taken over a suitably long interval of time, such as one year, to eliminate various seasonal complications.

*IDV* has the useful property of being independent of solar wind speed and is highly correlated with the near-Earth Heliospheric Magnetic Field (HMF) strength *B*. Thus once *IDV* is determined, solar wind *B* is known as well. *Svalgaard and Cliver* [2005] used *IDV* augmented with *Bartels*' [1932] *u*-measure to reconstruct the HMF strength for the years 1872-2004.

Here we report on an extension of the *IDV* index for a longer time interval (1835-2009), using many more stations. The inclusion of more data is particularly important for the years from 1890-1909 for which the initial version of the index (IDV05) was based on observations from only one station before 1901 and four more stations from 1903. An important aspect of IDV09 is that it includes recent years with index values at the same level as the very low values in 1901-1902, thus allowing the correlation between *IDV* and the magnitude of the near Earth HMF to be extended to such low values without extrapolation. With this correlation, we infer HMF *B* for years prior to the space age and compare our *B* values with those obtained by other investigators using geomagnetic or cosmic ray data.

## 2. Analysis

### 2.1 Derivation of IDV09

Our determination of IDV09 is essentially identical to that of IDV05 except for the inclusion of more data. In *Svalgaard and Cliver* [2005] we emphasized that *IDV* is a modern version of the *u*-measure building on ideas of a century ago [*Moos*, 1910]. *Kertz* [1958], *Mayaud* [1980], and *Svalgaard* [2005] suggested using only night-time values to avoid contamination by the regular diurnal variation, $S_R$. We followed their lead but further limit the time interval to only one hour following local midnight and constructed the interdiurnal variability index (*IDV*) for a given station as the unsigned difference between two consecutive days of the average value over the interval of a *H* component measured in nT. The individual unsigned differences were then averaged over longer intervals, *e.g.*, one full year (minimizing various geometric and seasonal effects, *e.g.* the semiannual non-solar variation due to the tilt of the Earth's dipole – a plot of 27-day Bartels Rotation values of *IDV* can be found in Svalgaard [2009]).

Since 2005, we have been collecting, digitizing, quality controlling, and correcting (where needed) hourly historical geomagnetic data from individual observatories as well as from World Data Centers [there is, as yet, no mechanism for injecting new or corrected data into the World Data Centers or various National Depositories, so we offer the data to interested researchers upon request]. Here we use these newly-acquired data to substantiate the *IDV*-index, which is especially important for the first ~30 years of the time series (1872-1902), during which IDV05 was based solely on *Schmidt's* [1926] and *Bartels'* [1932] *u*-measure from 1872-1889, on Potsdam observations from 1890-1902, plus Cheltenham for 1901-1902, and Honolulu for 1902. In contrast, IDV09 is based on four times as many "station years" (143 vs. 34) for this 31-yr interval as detailed in Table E1 of the Electronic Supplement. We update the time series by adding the index values for 2004-2009. These latter years are significant because the yearly-averages of *B* observed in 2007-2009 are the lowest observed during the space age. They lie at the lower endpoint of the correlation between yearly averages of observed *B* and *IDV*.

Table 1 contains a list of the 71 stations (including replacement stations) used to compute IDV09 (versus 14 for IDV05). A comprehensive list of the data coverage and the data values for the individual stations used in this study is given in Table E1 in the Electronic Supplement. All raw data is available from the authors (LS) upon request.

### 2.1.1. Latitude Normalization

For IDV05, we normalized *IDV* values for a given station with Corrected Geomagnetic Latitude, *M*, to those of Niemegk (NGK) [as Bartels did for the *u*-measure] using

$$IDV_{norm} = IDV_{raw} / (1.324 \cos^{0.7}(M)) \qquad (1)$$

Here we have retained this relationship for stations with $|M| < 51°$ because it still fits the data for the additional stations. At significantly higher latitudes, the index becomes strongly contaminated by auroral zone activity [see Figure 2 of *Svalgaard and Cliver*, 2005, and we recommended not using such stations, *e.g.*, the long-running station Sodankylä, SOD (used by *Lockwood et al.* [2009]). For IDV09, we relax this restriction slightly [by a few degrees for a few stations, indicated in Table 1] using a constant, empirical normalization divisor of 1.1 instead of the divisor in equation (1). A value 87  larger than ~0.95 for $|M| \geq 51°$ indicates some contamination by auroral zone activity.
88  We have not attempted to further quantify the latitude dependence of the contamination,
89  but simply use an average value for the few stations slightly above 51°. We do this to
90  accommodate changes in $M$ with time which for some stations can exceed several
91  degrees[1] and to include a few long-running stations just above 51°. Figure 1 shows the
92  adopted normalization divisor as a function of $M$ for the 71 stations used in the present
93  study. Different symbols denote the divisor values for the years 1800, 1900, and 2000,
94  showing the sensitivity of *IDV* to changes in latitude. The normalization divisor was
95  calculated for the centroid of the latitudes for the actual data coverage for each station. If
96  we did not normalize, the presence of data gaps [of which there are many] would produce
97  discontinuities in the composite series.

### 2.1.2. Effect of Hourly Means versus Hourly Values on *IDV*

99   Early magnetometer data were taken [and/or reported] as readings once an hour rather
100  than as the hourly mean that Adolf Schmidt advocated in 1905 and that was widely and
101  rapidly adopted. In *Svalgaard & Cliver* [2005] we showed that although the variance of
102  single values is larger than for averages, the overall effect on *IDV* was small (at most a
103  few percent) [2]. The two long-running series POT-SED-NGK and PSM-VLJ-CLF afford a
104  convenient additional test of this: POT changed from values to means with the 1905
105  yearbook, but CLF changed much later, with the 1972 yearbook, so we can directly
106  compare the (raw – uncorrected in any way) *IDV*-values for the two series (Figure 2). It is
107  evident that the change from hourly instantaneous values to hourly means did not
108  introduce any sudden changes in *IDV* at the times of the transitions. The Japanese station
109  at KAK changed from values to means in 1955. The ratio between raw *IDV* for KAK and
110  SED-NGK (crosses on Figure 2) also does not show any change in 1955. The American
111  stations CLH and HON changed to means with the 1915 yearbook. Comparison [Figure
112  3] over a 24-year interval centered on 1915 with the stations VLJ and DBN, which did
113  not change sampling procedure, also shows no detectable change in *IDV* due to the
114  change in sampling: the ratio between average CLH-HON and VLJ-DBN is 1.0792
115  before 1915 and 1.0792 after the change to hourly means in 1915. We conclude that
116  changes are too small to justify attempting *ad-hoc* correction based on extrapolation of
117  modern data.

### 2.1.3. Using the *u*-measure before 1872

119  Julius *Bartels* [1932] compiled the *u*-measure from the interdiurnal variability of the
120  Horizontal Component, $H$, from hourly or daily values from several observatories
121  operating from 1872 onwards as described in his paper. He wrote, "Before 1872, no
122  satisfactory data for the calculation of interdiurnal variabilities are available", but "more
123  for illustration than for actual use", he attempted to extend the series backwards to 1835.
124  For this he used the "Einheitliche Deklinations-Variationen"[3], $E$, of *Wolf* [1884] and the

---

[1] We expect only a very *weak* influence in the basic response of the Ring Current [see section 2.1.5] to the change of the Earth's magnetic dipole moment [as per *Glassmeier et al.*, 2004] over the interval in question, and so have not attempted to correct for this.
[2] This effect is significant for the *IHV* index, but for that case, correction of the effect is straightforward [*Svalgaard and Cliver*, 2007b].
[3] Unified Declination Variations

125  "summed ranges", $s$, derived from the mean diurnal variation of $H$ at Colaba (Bombay)
126  due to *Moos* [1910]. He derived regression formulae relating $E$ and $s$ to $u$ for times after
127  1872 and used them to synthesize values of $u$ for the earlier years[4]; giving $s$ double the
128  weight of $E$. Bartels justified this by showing that for the annual means 1872-1901, the
129  values of $u$ derived from $H$ and the values of $s$ have a high linear correlation coefficient.
130  We have extended his analysis by calculating the correlation between $IDV$ and the
131  Summed Ranges for 1872-1905 [Figure 4, top panel] finding a correlation coefficient of
132  0.86. Figure 4 [bottom panel] shows the agreement between observed $IDV$ [red] and that
133  calculated from $s$ [blue].

134  Furthermore, as shown in *Svalgaard and Cliver* [2005] there is a good linear correlation
135  between $IDV$ [or HMF $B$ derived from it] and the square root of the sunspot number, $R$.
136  The main sources of the equatorial components of the Sun's large-scale magnetic field
137  are large active regions. If these active regions emerge at random longitudes, their net
138  equatorial dipole moment will scale as the square root of their number. Thus their
139  contribution to the average HMF strength will tend to increase as $R^{1/2}$ (for a detailed
140  discussion, see *Wang and Sheeley* [2003] and *Wang et al.* [2005]).

141  To the extent that the $u$-measure before 1872 can be taken as a geomagnetic-based
142  measure of the sunspot number, it is therefore to be expected that the $u$-measure also will
143  serve as a proxy for $IDV$. This estimate will be independent of any assumptions about the
144  constancy of the calibration of the sunspot number (*c.f.* the difference between the Zürich
145  Sunspot Number and the Group Sunspot Number [*Hoyt et al.*, 1994]).

146  Figure 5 shows that $IDV$ can also be directly inferred from the daily range, $rY$, of the East
147  component [equivalent to the Declination for this purpose] of the geomagnetic field and
148  that therefore, again, that the $u$-index before 1872 [strongly influenced by the range of the
149  daily variation] can be used for estimation of $IDV$, albeit with slightly less accuracy than
150  after 1872. This conclusion may seem at variance [and did surprise us] with our initial
151  decision to use only night-time data in the derivation of $IDV$, but emerges naturally [and
152  inescapably] after our analysis had shown that $IDV$ derived without any dependence on
153  daytime data is comparable to $IDV$ derived from daily ranges because of the strong
154  dependence of both on the sunspot number. This is clearly demonstrated in Figure 6 that
155  shows raw $IDV$ calculated for PSM-VLJ-CLF and POT-SED-NGK determined from
156  night-time differences (blue) and daytime differences (red). This realization opens the
157  door for use of 19[th] geomagnetic stations that only observed during the day as long as the
158  observations were made at fixed hours.

159  For the reasons given above, we find that $IDV$ can be estimated with confidence from
160  Bartels' $u$-measure also before 1872, justifying our reconstruction of HMF $B$ since 1835.

161  **2.1.4. The *IDV*-index 1835-2009**

162  From the ~1,375,000 daily differences [3775 station-years] derived from the stations in
163  Table 1 we construct the $IDV$-index shown in Figure 7, with individual station curves in
164  grey. The composite (red) curve is the mean of the median and average values for each
165  year, while before 1872 the dashed curve shows $IDV$ estimated from $u$. Also shown (blue

---

[4] From $E$ and $s$, we calculate a value of 0.72 for the value for $u$ for 1857 using the formulae given by Bartels.

166  curve) is the number of stations contributing to the mean. The large number of stations
167  from 1957 on does not add further significance to the composite, but only serves to
168  establish the range of scatter of the values.

169  It is evident that *IDV* from only a single station (provided that not too much data is
170  missing either because the recording went off-scale or as a result of other problems) does
171  not differ much from the mean of many stations; the standard deviation of *IDV*-values for
172  all stations for a given year is less than 1 nT or about 9%. This means that only a few
173  [good] stations are needed for a robust determination of *IDV*. This conclusion, of course,
174  only emerges after the spread of *IDV*-values has first been shown to be small. The
175  standard error of the mean of more than fifty stations is 0.1 nT.

176  Figure 8 shows that the differences between IDV05 and IDV09 are slight, and due to the
177  additional data since 1880. During the period of overlap (1872-2003, 2004 was only
178  partial), the two time series agree within an RMS of 0.33 nT or 3%. The coefficient of
179  determination for the correlation between IDV09 and IDV05 is $R^2 = 0.984$. *IDV* is a
180  robust index.

### 2.1.5. Physical Interpretation of *IDV*: Measure of the Energy in the Ring Current

181
182  In *Svalgaard and Cliver* [2005] we reported that *IDV* is closely correlated with the
183  negative part of the $D_{st}$-index based on data back to 1932 [*Karinen and Mursula*, 2005].
184  In *Svalgaard and Cliver* [2006] we extended that relationship back to 1905 using the 100-
185  year $D_{st}$-series derived by *J. Love* [2006, 2007], and confirm it here using IDV09. Yearly
186  averages of $D_{st}$ [scaled to Kyoto $D_{st}$; we use $D_{st}$ here in a generic sense without
187  distinguishing between different derivations of the underlying physical measure sought
188  captured by $D_{st}$] when the hourly value was negative were computed and found to be
189  strongly correlated with *IDV* [$R^2 = 0.91$]: $IDV = -0.45$ ($D_{st}<0$). Figure 9 compares IDV09
190  and *IDV* computed from $D_{st}$. The good match suggests that *IDV* is a measure of the same
191  physical reality as negative $D_{st}$, namely the energy in the Ring Current, which then in turn
192  seems to be controlled by HMF *B*: ($D_{st}<0$) = 4.81 $B$ − 9.41 [$R^2 = 0.84$], and we can then
193  also use $D_{st}$ to determine the HMF strength: $B = 2.70 - 0.1736$ ($D_{st}<0$). *Schmidt* [1926]
194  actually suggested that in the definition of the *u*-measure it would be slightly better to
195  only use the negative differences between consecutive days.

### 2.2. Using IDV09 to Calculate HMF Strength, 1835-2009

196
197  Since the 2005 definition paper, lower values of HMF strength, *B*, have improved the
198  dynamic range (and thus the statistical significance) of the correlation between *IDV* and
199  *B*. An approximate linear correlation was found, but there is no *a priori* reason the
200  relationship would be strictly linear. In addition, it has been argued [*Lockwood et al.*
201  2006] that *B* should be taken as the independent variable instead of *IDV*. We showed in
202  *Svalgaard and Cliver* [2006] that it does not make much difference which way the
203  correlation is evaluated. In the end, the RMS difference [0.4 nT or less than ~10%]
204  between HMF *B* observed *in situ* near the Earth[5] and inferred from *IDV* is what matters.

---

[5] Using hourly averages from the OMNI dataset for historical data and from ACE for near real-time recent data.

The average coefficients for the linear correlation performed four ways (average, median, and for each: direct and inverse) are

$$B \text{ (nT)} = (2.06 \pm 0.21) + (0.441 \pm 0.021) \, IDV \quad (R^2 = 0.869) \quad (2)$$

The equivalent power law dependence comes to

$$B \text{ (nT)} = (1.33 \pm 0.07) \, IDV^{\,0.689 \pm 0.023} \quad (R^2 = 0.905) \quad (3)$$

The adopted values for $B$ inferred from IDV09 given in Table 2 are the mean values calculated using these two relationships. Table E3 in the Electronic Supplement summarizes the coefficients for all correlations. The 'error bars' quoted are not a measure of the statistical significance of the correlations in a strict sense, but are solely indicative of the range or variability of the various coefficients.

Figure 10 shows the values for HMF $B$ inferred from $IDV$ from 1835 to the present (blue curve) and $B$ measured by spacecraft (red curve). A 4th-order polynomial fit suggests a ~100 year Gleissberg cycle. Cycle 23 looks remarkably like cycle 13, including the very deep solar minimum following both cycles, likely presaging a weak cycle 24 as predicted from the solar polar fields [*Svalgaard et al.*, 2005; *Schatten*, 2005]. It is clear that we are returning to conditions prevailing a century ago. It seems likely that other solar parameters such as Total Solar Irradiance [*Fröhlich*, 2009; *Steinhilber et al.*, 2009] and cosmic ray modulation [*Steinhilber et al.*, 2010] are reverting to similar conditions with possible implications for the climate-change debate.

## 2.3. Comparison of IDV09-based $B$ with Other Recent Reconstructions

## 2.3.1. Consilience of Reconstructions Based on Geomagnetic Data.

Reconstructions of HMF $B$ have been discordant in the past [e.g. *Schatten et al.*, 1978; *Andreasen*, 1997; *Lockwood et al.*, 1999, 2006; *Svalgaard and Cliver*, 2005, 2006, 2007b]. The realization [*Svalgaard et al.*, 2003] that geomagnetic indices can be constructed that have different dependencies on $B$ and solar wind speed ($V$) has enabled robust determinations of both $V$ [*Svalgaard and Cliver*, 2007b; *Rouillard et al.*, 2007; *Lockwood et al.*, 2009] and $B$ [*Svalgaard and Cliver*, 2005, 2006; *Lockwood et al.*, 2009] that have converged to a common, well-constrained dataset. Progress has been swift and Figure 11 shows the convergence of HMF $B$ determined by *Lockwood et al.* [2009] to the values determined from $IDV$ [*Svalgaard and Cliver*, 2005, this paper]. The *Lockwood et al.* [2009, and references therein] reconstruction still differs from ours for a few years during solar cycle 14, but apart from that, the agreement is quite remarkable and the issues seem resolved.

Figure 12 details the evolution of the various determinations of $B$ since the seminal, but now superseded, *Lockwood et al.* [1999] paper. It is clear that we now possess the methodology to infer B with good accuracy as far back as continuous geomagnetic records of $H$ reach. A concerted effort of digitization of 19th century yearbook records would promise to further improve our knowledge of the magnetic field in the heliosphere.

*Svalgaard and Cliver* [2007a] argued for a floor in yearly averages of solar wind $B$ which was approached at every 11-yr minimum and represented the ground-state of the Sun during extended minima such as the Maunder Minimum. With the larger dynamic range afforded by the current minimum, we can now refine the value of the floor to be closer to

247    the ~4 nT observed during 2008 and 2009 [see also *Owens et al.*, 2008], returning to the
248    values inferred for 11-yr minima during the previous Gleissberg minimum at the turn of
249    the 20th century.

## 250    2.3.2. Comparison with [10]Be-based Reconstructions

251    *McCracken* [2007] spliced together [10]Be data, ionization-chamber cosmic ray data
252    (calibrated with balloon flight data), and neutron monitor cosmic ray data to produce an
253    'equivalent' neutron monitor count series covering the entire interval 1428-2005, and
254    inverted the series for *B* in order to express the data in terms of the HMF *B*. In Figure 13
255    we compare his series for HMF *B* with the 'consensus' *B* from geomagnetic data.

256    In McCracken's time series for *B*, a large step-like change (1.7 nT; from 3.5 nT to 5.2 nT;
257    the largest jump in the entire ~600-year record) occurs between the 1944 and 1954
258    sunspot minima flanking cycle 18. No such corresponding change is observed in the
259    concordant reconstructions of *Svalgaard and Cliver* [2005; this paper], *Rouillard et al.*
260    [2007] and *Lockwood et al.* [2009], nor in *B* calculated from the quantity *BV* deduced by
261    *Le Sager and Svalgaard* [2004] using either *V* of *Svalgaard and Cliver* [2006] or of
262    *Rouillard et al.* [2007], or in *B* deduced from $D_{st}$.

263    *Muscheler et al.* [2007] discuss the uncertainties with the balloon-borne data that form
264    the basis for McCracken's calibration of the composite equivalent neutron monitor data
265    before 1951. The strong geomagnetic evidence argues that the calibration of the pre-
266    neutron monitor cosmic ray reconstruction is not on a firm footing. We suggest that part
267    of the reason for the disagreement might lie with the calibration and splicing together of
268    the disparate cosmic ray datasets.

269    After our paper was submitted, we were pleased to read a paper by *Steinhilber et al.*
270    [2010] in which a new [10]Be-based reconstruction has moved closer to our reconstruction,
271    to that of *Rouillard et al.* [2007], and to that of *Caballero-Lopez et al.* [2004] with
272    diffusion coefficient depending inversely of B$^2$ (a ~ 2). The reconstruction of *Steinhilber*
273    *et al.* [2010] still differs somewhat with the geomagnetic based reconstructions,
274    especially for the ~1880-1900 interval [Figure 14] and, just like the previous discrepancy,
275    this will need to be resolved. We suggest that if the sharp dip around ~1895 is not borne
276    out by further investigation, the magnitude of earlier excursions to very low values may
277    also be in doubt. Figure 15 shows *IDV* for all stations for the interval 1880-1920 and does
278    not support the marked decrease around ~1895. It is unlikely that further stations will
279    change that conclusion.

## 280    3. Summary and Discussion

281    We have extended our 1872-2004 HMF time series [*Svalgaard and Cliver*, 2005] to the
282    years 1835-2009 [Figure 10]. The 1835-1871 interval is based on Bartels' *u*-measure,
283    which he extended from 1871 back to 1835 using *Wolf*'s [1884] Declination index based
284    on several European stations and *Moos*' [1910] summed ranges from Colaba. The 1872-
285    2009 interval is based on the *IDV*-index, with significantly more data for the early years
286    (1872-1910). The forward extension of the HMF series through 2009 is important
287    because the years 2007-2009 witnessed the lowest annual averages of *IDV* during the
288    space age. For the time of overlap between the re-evaluated *IDV*-index (IDV09) and
289    IDV05, the difference is very small, testifying to the robustness of the index.

A comparison of IDV09-based HMF strength with those obtained by other investigators using various combinations and permutations of geomagnetic indices revealed a pleasing agreement [Figure 11] in what had been previously a contentious field of research [Figures 12 and 13]. The technique proposed by *Svalgaard et al.* [2003] and adopted by *Rouillard et al.* [2007] to use indices with different dependencies on $B$ and $V$ to separate these variables has proven out and allowed the vast storehouse of hourly and daily data to be brought to bear. In particular, the $B$ values deduced and cross-checked [*Le Sager and Svalgaard*, 2004] by this method have substantiated the approach made possible by the *IDV*-index and, as we suggested in *Svalgaard and Cliver* [2005], and have confirmed here, the negative component of the $D_{st}$-index [Figure 9]. We conclude that the long-term variation of heliospheric $B$ is firmly constrained during the time for which it is based on hourly values of $H$, and that current values at the solar minimum between cycles 23 and 24 are back to where they were 108 years ago at the solar minimum between cycles 13 and 14.

Although the recent reconstruction of $B$ based on $^{10}$Be data [*Steinhilber et al.*, 2010] generally agrees well with the geomagnetic-based reconstruction there is disagreement for the decade just prior to 1900 [Figure 14]. Further examination of these years is critical because they present the only example during the 175 year interval of geomagnetic-based $B$ where the floor [*Svalgaard and Cliver*, 2007a] in the solar wind is challenged by *Steinhilber et al* [2010], but not actually supported by the geomagnetic evidence [Figure 15].

**Acknowledgements**

Geomagnetic data has been downloaded from the World Data Centers for Geomagnetism in Kyoto, Japan, and Copenhagen, Denmark [now defunct], and from INTERMAGNET at http://www.intermagnet.org/Data_e.html. The research results presented in this paper rely on the data collected at magnetic observatories worldwide, and we thank the national institutions that support them. We also recognize the role of the INTERMAGNET program in promoting high standards of magnetic observatory practice. We thank the many people worldwide who have helped us with collection of data and metadata in addition to what is available from public sources. We thank Vladimir Papitashvili for the program to calculate corrected geomagnetic coordinates using the GUFM1 coefficients (courtesy of Catherine Constable). The OMNI dataset was downloaded from http://omniweb.gsfc.nasa.gov/. Real-time ACE interplanetary data is downloaded from http://www.swpc.noaa.gov/ftpmenu/lists/ace2.html.

417   Table 1. Stations used for IDV09, including replacement stations due to relocation of
418   original stations. The Corrected Geomagnetic Latitude for the year 2000 is given for
419   illustration, but the centroid of the latitudes for the time of operation was used to estimate
420   the Normalization Constants. Constants in *italics* were determined by an empirical fit to
421   time-overlapping stations. For a few observatories (marked with an asterisk) weakly non-
422   linear relationships have been used to normalize directly to NGK. A list of IAGA
423   designations, observatory names, and other station details can be found at
424   http://www.geomag.bgs.ac.uk/gifs/annual_means.shtml.

| Stations (IAGA Abbrev.) | Geodetic Latitude | Geodetic Longitude | Corrected Geomagnetic Latitude 2000 | Divisor |
|---|---|---|---|---|
| BOX | 58.0 | 39.0 | *53.9* | 1.10 |
| ESK* | 55.3 | 356.8 | *52.9* | *1.00* |
| EKT,SVD,ARS | 56.4 | 58.6 | *52.1* | 1.10 |
| RSV,BFE | 55.6 | 11.7 | *52.1* | 1.10 |
| MOS | 55.5 | 37.3 | *51.3* | 1.10 |
| NVS | 55.0 | 82.9 | 50.5 | 0.97 |
| WLH,WNG | 53.7 | 9.1 | 50.1 | 0.97 |
| MNK | 54.1 | 26.5 | 49.9 | 0.98 |
| CLH,FRD | 38.2 | 282.6 | 49.7 | 0.97 |
| BOU | 40.1 | 254.8 | 49.2 | 0.99 |
| BAL | 38.8 | 264.8 | 49.0 | 0.99 |
| DBN,WIT | 52.1 | 5.2 | 48.4 | 0.98 |
| 10u | 52.4 | 13.1 | 48.3 | 1.00 |
| POT,SED,NGK | 52.1 | 12.7 | 48.0 | 1.00 |
| ABN,HAD | 51.0 | 355.5 | 47.8 | 0.99 |
| BEL | 51.8 | 20.8 | 47.5 | 1.01 |
| IRT | 52.2 | 104.5 | 47.0 | 1.02 |
| TKT | 41.3 | 69.6 | 46.5 | 1.08 |
| PET | 53.1 | 158.6 | 46.3 | 1.02 |
| DOU | 50.1 | 4.6 | 46.0 | 1.02 |
| LVV | 49.9 | 23.8 | 45.3 | 1.04 |
| PSM,VLJ,CLF | 48.0 | 2.3 | 43.6 | 1.04 |
| FUR | 48.2 | 11.3 | 43.4 | 1.05 |
| HRB | 47.9 | 18.2 | 43.0 | 1.06 |
| THY | 46.9 | 17.9 | 41.8 | 1.08 |
| YSS | 47.0 | 142.7 | 39.9 | 1.10 |

| | | | |
|---|---|---|---|
| TUC | 32.3 | 249.2 | 39.9 | 1.10 |
| AAA | 43.3 | 76.9 | 38.4 | 1.12 |
| TFS | 42.1 | 44.7 | 37.2 | 1.14 |
| MMB | 43.9 | 144.2 | 36.7 | 1.13 |
| AQU | 42.4 | 13.3 | 36.3 | 1.13 |
| BJI,BMT | 40.3 | 116.2 | 34.2 | 1.16 |
| SFS,EBR | 40.8 | 0.5 | 34.2 | 1.14 |
| COI | 40.2 | 351.6 | 34.1 | 1.15 |
| LNP,LZH | 36.1 | 103.9 | 30.1 | 1.20 |
| VQS,SJG | 18.4 | 293.9 | 29.2 | 1.20 |
| KAK | 36.2 | 140.2 | 28.9 | 1.20 |
| KNZ | 35.3 | 140.0 | 27.9 | 1.21 |
| HTY | 33.1 | 139.8 | 25.7 | 1.23 |
| SSH | 31.1 | 121.2 | 24.4 | 1.24 |
| KNY | 31.4 | 130.9 | 24.3 | 1.24 |
| HON | 21.3 | 202.0 | 21.7 | 1.26 |
| GUI | 28.3 | 343.6 | 15.7 | 1.29 |
| PHU | 21.0 | 106.0 | 13.7 | 1.30 |
| API | 13.8 | 188.2 | 12.8 | 1.30 |
| ABG | 18.6 | 72.9 | 11.8 | 1.31 |
| KOU | 5.1 | 307.4 | 10.8 | 1.30 |
| MBO | 14.4 | 343.0 | 3.2 | 1.31 |
| ANN | 11.4 | 79.7 | 3.1 | 1.32 |
| TAM | 22.8 | 5.5 | 3.1 | 1.32 |
| HUA | -12.1 | 284.7 | 2.1 | 1.32 |
| GUA | 13.6 | 144.9 | 1.0 | 1.32 |
| TRD | 8.5 | 77.0 | 0.4 | 1.32 |
| AAE | 9.0 | 38.8 | -1.3 | 1.32 |
| BNG | 4.4 | 18.6 | -2.2 | 1.32 |
| ASC | -7.5 | 345.6 | -7.9 | 1.32 |
| BTV | -6.2 | 106.8 | -15.8 | 1.29 |
| PPT | -17.6 | 210.4 | -16.4 | 1.29 |
| VSS | -22.4 | 316.4 | -16.5 | 1.30 |
| PIL | -31.7 | 296.1 | -18.6 | 1.28 |
| TAN | -18.9 | 47.6 | -29.1 | 1.20 |
| TSU | -19.2 | 17.7 | -30.0 | 1.20 |
| HBK | -22.9 | 27.7 | -33.6 | 1.17 |
| CTO,HER | -34.4 | 19.2 | -42.3 | 1.09 |
| WAT,GNA | -31.8 | 116.0 | -44.4 | 1.05 |
| TOO,CNB | -35.3 | 149.4 | -45.8 | 1.04 |
| TRW | -43.3 | 19.0 | -47.8 | 1.02 |
| AMS* | -37.8 | 77.6 | -49.1 | 1.00 |
| AIA | -65.2 | 295.7 | -49.8 | *1.20* |
| AML,EYR | -43.4 | 172.4 | -50.3 | 0.97 |
| CZT | -46.4 | 51.9 | *-53.1* | 1.10 |

Table 2. IDV09: The *IDV*-index [measured or inferred] for each year since 1835. The HMF strength *B* at the Earth is derived from *IDV* as per section 2.2. The field observed *in situ* [OMNI/ACE datasets] is given for comparison. A few years had very incomplete data coverage and missing data were derived by linear interpolation across data gaps to avoid uneven coverage skewing the average. Those values are in *italics*.

| Year | IDV09 | IDV HMF *B* | Obs HMF *B* |
|---|---|---|---|
| 1835.5 | 11.60 | 7.23 | |
| 1836.5 | 16.30 | 9.30 | |
| 1837.5 | 16.00 | 9.17 | |
| 1838.5 | 16.80 | 9.51 | |
| 1839.5 | 14.00 | 8.30 | |
| 1840.5 | 12.20 | 7.50 | |
| 1841.5 | 10.10 | 6.55 | |
| 1842.5 | 9.00 | 6.05 | |
| 1843.5 | 8.90 | 6.00 | |
| 1844.5 | 8.50 | 5.81 | |
| 1845.5 | 9.50 | 6.28 | |
| 1846.5 | 10.10 | 6.55 | |
| 1847.5 | 11.40 | 7.14 | |
| 1848.5 | 13.10 | 7.90 | |
| 1849.5 | 12.00 | 7.41 | |
| 1850.5 | 9.90 | 6.46 | |
| 1851.5 | 9.00 | 6.05 | |
| 1852.5 | 7.60 | 5.39 | |
| 1853.5 | 7.80 | 5.48 | |
| 1854.5 | 7.60 | 5.39 | |
| 1855.5 | 5.80 | 4.51 | |
| 1856.5 | 6.60 | 4.90 | |
| 1857.5 | 7.20 | 5.19 | |
| 1858.5 | 10.60 | 6.78 | |
| 1859.5 | 14.30 | 8.43 | |
| 1860.5 | 13.50 | 8.08 | |
| 1861.5 | 12.20 | 7.50 | |
| 1862.5 | 10.00 | 6.51 | |
| 1863.5 | 9.40 | 6.23 | |
| 1864.5 | 8.40 | 5.76 | |
| 1865.5 | 7.90 | 5.53 | |
| 1866.5 | 7.30 | 5.24 | |
| 1867.5 | 6.80 | 5.00 | |
| 1868.5 | 8.10 | 5.62 | |
| 1869.5 | 11.10 | 7.01 | |
| 1870.5 | 16.70 | 9.47 | |
| 1871.5 | 16.00 | 9.17 | |
| 1872.5 | 14.60 | 8.56 | |
| 1873.5 | 9.90 | 6.46 | |
| 1874.5 | 9.50 | 6.28 | |
| 1875.5 | 7.20 | 5.19 | |
| 1876.5 | 6.00 | 4.61 | |
| 1877.5 | 6.60 | 4.90 | |
| 1878.5 | 5.80 | 4.51 | |
| 1879.5 | 6.00 | 4.61 | |
| 1880.5 | 7.89 | 5.53 | |
| 1881.5 | 8.69 | 5.90 | |
| 1882.5 | 12.47 | 7.62 | |
| 1883.5 | 10.20 | 6.60 | |
| 1884.5 | 9.36 | 6.22 | |
| 1885.5 | 9.57 | 6.22 | |
| 1886.5 | 9.14 | 6.11 | |
| 1887.5 | 7.75 | 5.46 | |
| 1888.5 | 7.27 | 5.23 | |
| 1889.5 | 6.99 | 5.09 | |
| 1890.5 | 6.22 | 4.72 | |
| 1891.5 | 8.60 | 5.86 | |
| 1892.5 | 14.02 | 8.31 | |
| 1893.5 | 10.79 | 6.87 | |
| 1894.5 | 13.12 | 7.91 | |
| 1895.5 | 9.95 | 6.48 | |
| 1896.5 | 10.48 | 6.73 | |
| 1897.5 | 8.71 | 5.91 | |
| 1898.5 | 8.98 | 6.04 | |
| 1899.5 | 7.06 | 5.13 | |
| 1900.5 | 5.75 | 4.49 | |
| 1901.5 | 4.90 | 4.06 | |
| 1902.5 | 5.04 | 4.13 | |
| 1903.5 | 7.03 | 5.11 | |
| 1904.5 | 7.54 | 5.36 | |
| 1905.5 | 8.62 | 5.87 | |
| 1906.5 | 7.49 | 5.33 | |
| 1907.5 | 8.83 | 5.97 | |
| 1908.5 | 9.45 | 6.26 | |
| 1909.5 | 9.95 | 6.48 | |
| 1910.5 | 8.10 | 5.63 | |
| 1911.5 | 7.08 | 5.14 | |
| 1912.5 | 5.69 | 4.46 | |
| 1913.5 | 5.15 | 4.18 | |
| 1914.5 | 6.22 | 4.72 | |
| 1915.5 | 8.09 | 5.62 | |
| 1916.5 | 9.19 | 6.13 | |
| 1917.5 | 10.95 | 6.94 | |
| 1918.5 | 10.97 | 6.95 | |

| Year | | | | Year | | | |
|---|---|---|---|---|---|---|---|
| 1919.5 | 11.57 | 7.22 | | 1965.5 | 6.93 | 5.07 | 5.06 |
| 1920.5 | 10.28 | 6.64 | | 1966.5 | 7.88 | 5.52 | *6.00* |
| 1921.5 | 8.97 | 6.03 | | 1967.5 | 10.30 | 6.65 | 6.36 |
| 1922.5 | 7.74 | 5.45 | | 1968.5 | 9.47 | 6.26 | 6.19 |
| 1923.5 | 6.17 | 4.70 | | 1969.5 | 9.39 | 6.23 | 6.05 |
| 1924.5 | 6.89 | 5.04 | | 1970.5 | 10.12 | 6.56 | 6.35 |
| 1925.5 | 8.05 | 5.60 | | 1971.5 | 8.84 | 5.97 | 6.00 |
| 1926.5 | 10.69 | 6.82 | | 1972.5 | 9.49 | 6.27 | 6.38 |
| 1927.5 | 9.29 | 6.18 | | 1973.5 | 9.28 | 6.18 | 6.35 |
| 1928.5 | 9.69 | 6.37 | | 1974.5 | 9.18 | 6.13 | 6.63 |
| 1929.5 | 9.64 | 6.34 | | 1975.5 | 8.15 | 5.65 | 5.82 |
| 1930.5 | 10.22 | 6.61 | | 1976.5 | 8.70 | 5.91 | 5.45 |
| 1931.5 | 7.38 | 5.28 | | 1977.5 | 8.96 | 6.03 | 5.85 |
| 1932.5 | 7.22 | 5.21 | | 1978.5 | 12.32 | 7.56 | 7.08 |
| 1933.5 | 6.96 | 5.08 | | 1979.5 | 11.78 | 7.32 | 7.59 |
| 1934.5 | 6.83 | 5.02 | | 1980.5 | 10.51 | 6.74 | 6.98 |
| 1935.5 | 7.75 | 5.46 | | 1981.5 | 13.78 | 8.20 | 7.84 |
| 1936.5 | 8.81 | 5.96 | | 1982.5 | 15.25 | 8.84 | 8.81 |
| 1937.5 | 12.11 | 7.46 | | 1983.5 | 11.60 | 7.23 | *7.61* |
| 1938.5 | 14.02 | 8.31 | | 1984.5 | 10.50 | 6.74 | *7.32* |
| 1939.5 | 12.79 | 7.77 | | 1985.5 | 9.06 | 6.07 | 5.89 |
| 1940.5 | 12.48 | 7.63 | | 1986.5 | 8.80 | 5.95 | 5.74 |
| 1941.5 | 12.10 | 7.46 | | 1987.5 | 8.20 | 5.67 | 6.09 |
| 1942.5 | 9.57 | 6.31 | | 1988.5 | 10.21 | 6.61 | 7.30 |
| 1943.5 | 8.97 | 6.03 | | 1989.5 | 16.74 | 9.48 | 8.15 |
| 1944.5 | 8.28 | 5.71 | | 1990.5 | 12.84 | 7.79 | 7.29 |
| 1945.5 | 8.75 | 5.93 | | 1991.5 | 15.77 | 9.07 | 9.34 |
| 1946.5 | 14.33 | 8.44 | | 1992.5 | 12.87 | 7.80 | 8.25 |
| 1947.5 | 13.85 | 8.24 | | 1993.5 | 10.08 | 6.54 | 6.59 |
| 1948.5 | 10.87 | 6.91 | | 1994.5 | 9.06 | 6.07 | 6.15 |
| 1949.5 | 13.55 | 8.10 | | 1995.5 | 9.08 | 6.08 | 5.72 |
| 1950.5 | 12.56 | 7.66 | | 1996.5 | 6.76 | 4.98 | 5.11 |
| 1951.5 | 12.46 | 7.62 | | 1997.5 | 8.06 | 5.60 | 5.51 |
| 1952.5 | 10.97 | 6.95 | | 1998.5 | 10.34 | 6.66 | 6.89 |
| 1953.5 | 8.90 | 6.00 | | 1999.5 | 9.82 | 6.42 | 6.91 |
| 1954.5 | 7.46 | 5.32 | | 2000.5 | 13.36 | 8.02 | 7.18 |
| 1955.5 | 8.69 | 5.90 | | 2001.5 | 13.44 | 8.05 | 6.94 |
| 1956.5 | 13.38 | 8.02 | | 2002.5 | 10.90 | 6.92 | 7.64 |
| 1957.5 | 16.65 | 9.45 | | 2003.5 | 12.51 | 7.64 | 7.60 |
| 1958.5 | 15.42 | 8.92 | | 2004.5 | 9.42 | 6.24 | 6.53 |
| 1959.5 | 14.39 | 8.47 | | 2005.5 | 9.44 | 6.25 | 6.25 |
| 1960.5 | 15.87 | 9.11 | | 2006.5 | 7.22 | 5.21 | 5.03 |
| 1961.5 | 11.49 | 7.18 | | 2007.5 | 5.96 | 4.59 | 4.48 |
| 1962.5 | 8.62 | 5.87 | | 2008.5 | 5.29 | 4.25 | 4.23 |
| 1963.5 | 8.06 | 5.60 | 5.45 | 2009.5 | 5.04 | 4.13 | 4.05 |
| 1964.5 | 7.19 | 5.19 | 5.12 | *2010.2* | *5.50* | *4.45* | *4.95* |

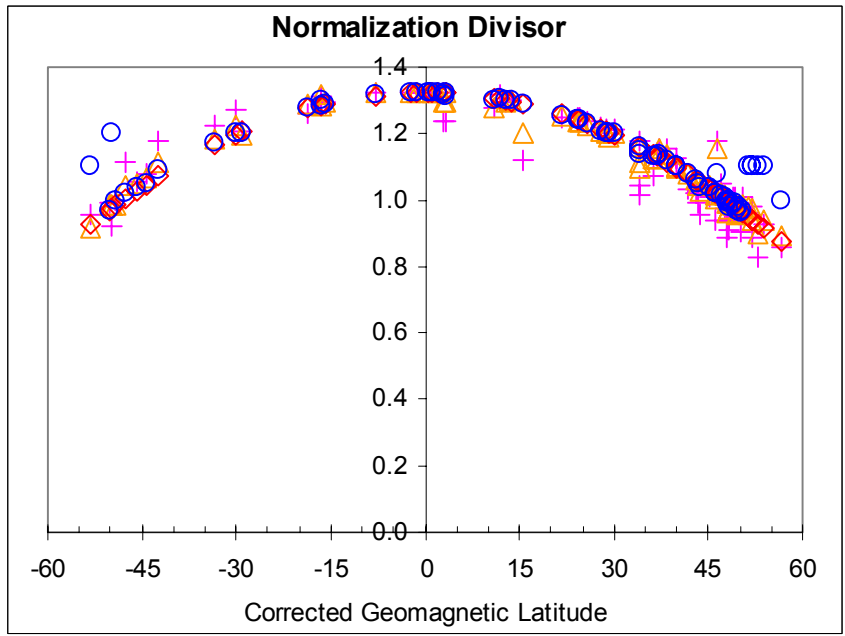

Figure 1. Adopted divisors (blue circles) to normalize *IDV* to the NGK-scale as a function of average corrected geomagnetic latitude for each station over the time of operation. For each station, different color coded symbols show what the divisor would have been for that station for years 1800 (pink pluses), 1900 (orange triangles), and 2000 (red diamonds).

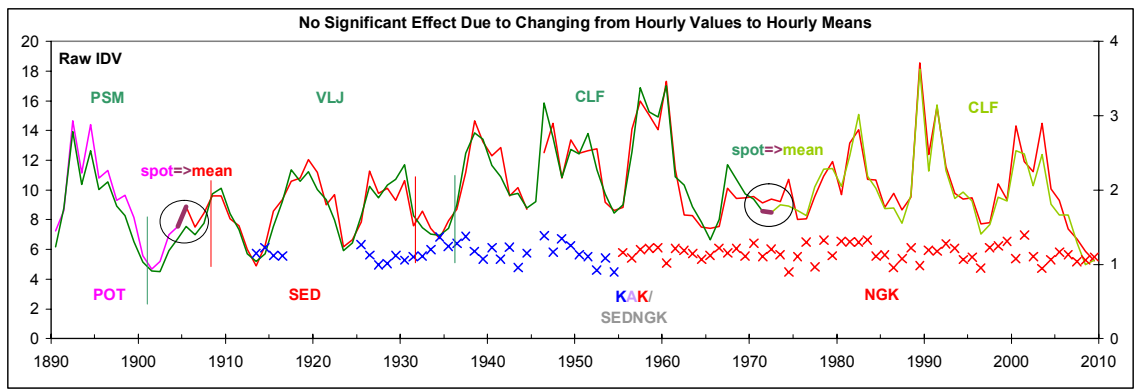

Figure 2. *IDV* calculated without any normalization or adjustments for the long-running German series (Potsdam POT–Seddin SED–Niemegk NGK; reddish curves) and the long-running French series (Parc Saint-Maur PSM–Val Joyeux VLJ–Chambon-la-Forêt CLF; greenish curves). Vertical lines show when the replacement stations went into operation and the ovals show when the yearbook values changed from being instantaneous hourly spot values to hourly means. The blue (spot values) and red (hourly means) crosses show the ratio between raw *IDV*s for KAK-Kakioka and SED-NGK. KAK changed from recording spot values to recording hourly means in 1955. There are no clear indications of changes in *IDV* due to the change in recording/reporting practice.

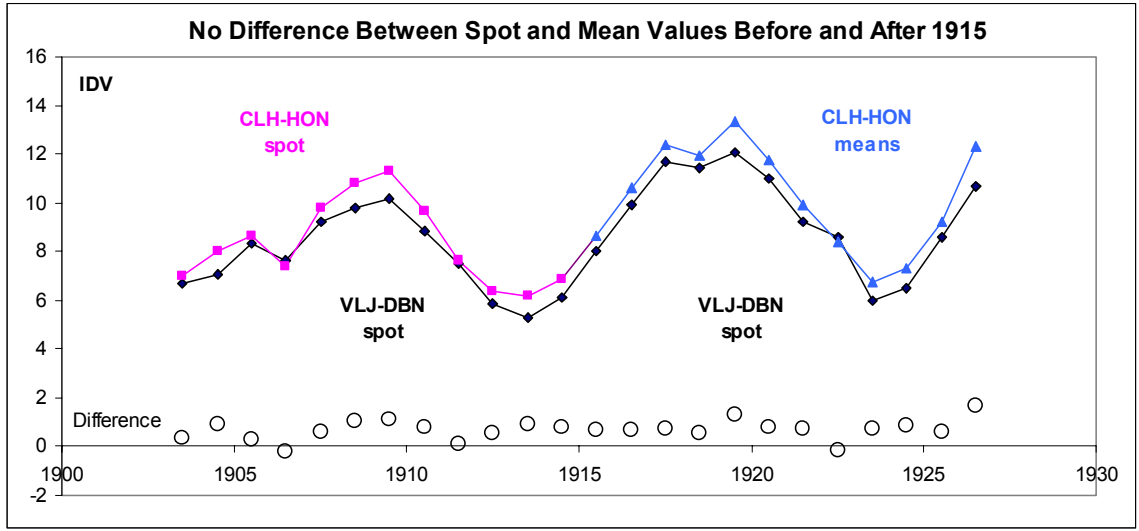

Figure 3. Raw *IDV* for the average of stations VLJ and DBN [both reporting instantaneous 'spot' values every hour for the 12-year intervals before 1915 and after 1915] and for the average of CLH and HON [reporting spot values before 1915 (pink) and hourly mean value thereafter (blue)].

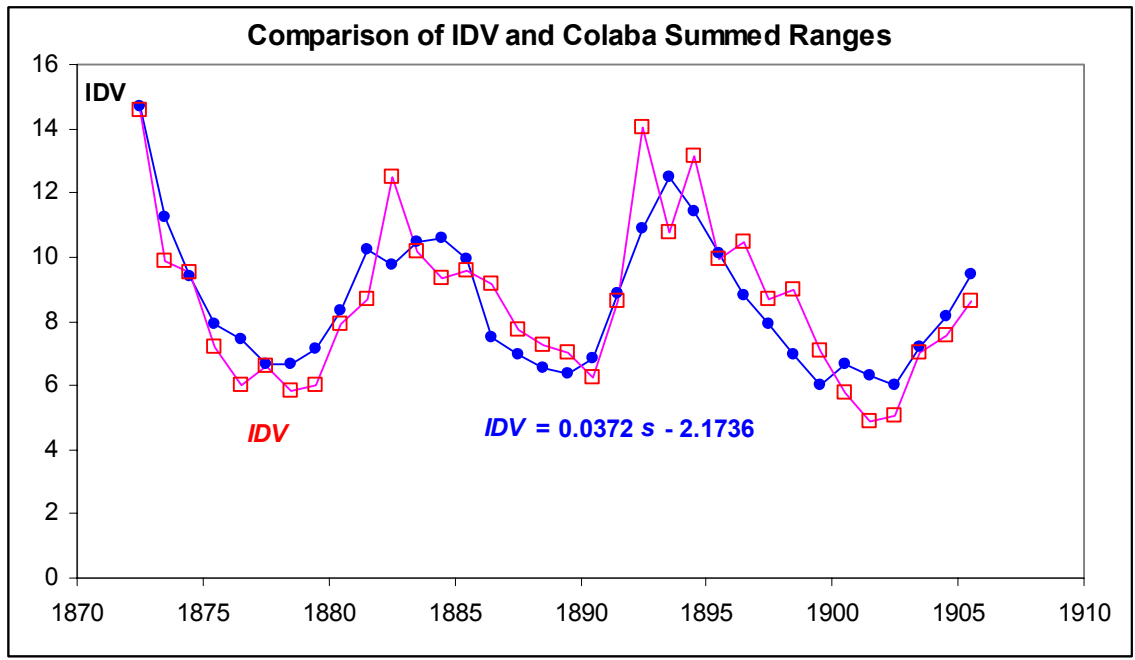

Figure 4 Comparison of observed *IDV* (red open squares) and synthetic *IDV* calculated from *s* using the regression equation shown (filled blue circles) derived from the correlation between *IDV* and the Summed Ranges, *s*, of *H* from Colaba [*Moos*, 1910; page 294, table 261] for 1872-1905.

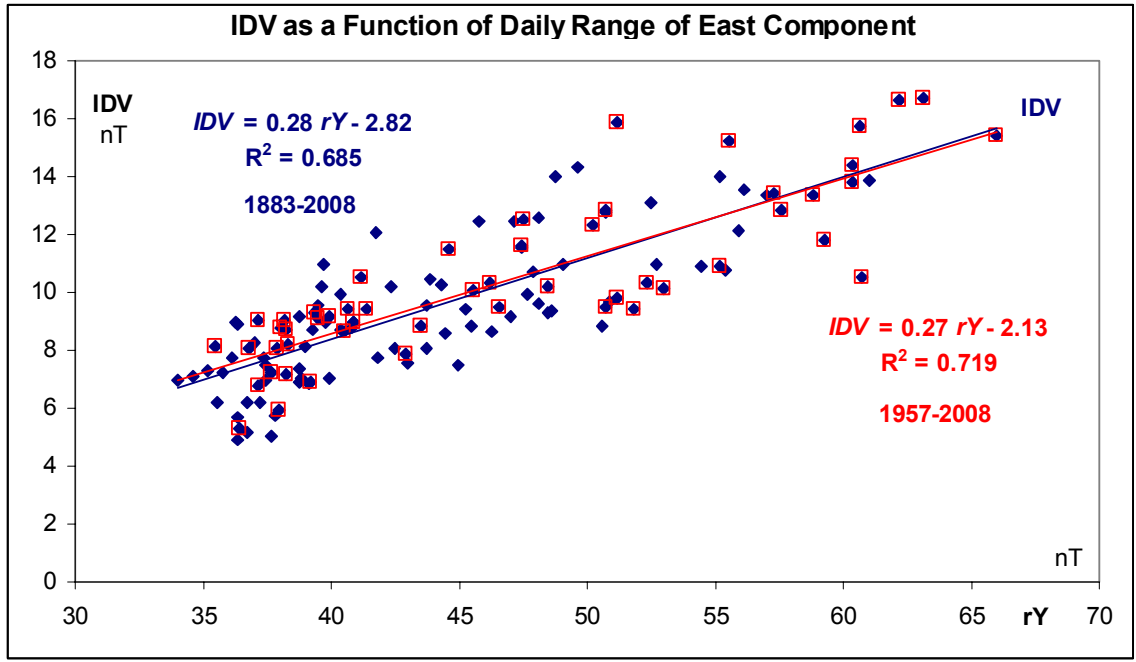

Figure 5. *IDV* plotted against the amplitude of the daily range, *rY*, of the East component [Table E2] of the geomagnetic field derived from PSM-VLJ-CLF and POT-SED-NGK, covering the interval 1883-2008 [dark blue diamonds] for which we have data for these stations. Since 1957, the number of stations contributing to *IDV* is high [~50] for every year, so the data is good. The open red squares show the same relationship for 1957-2008.

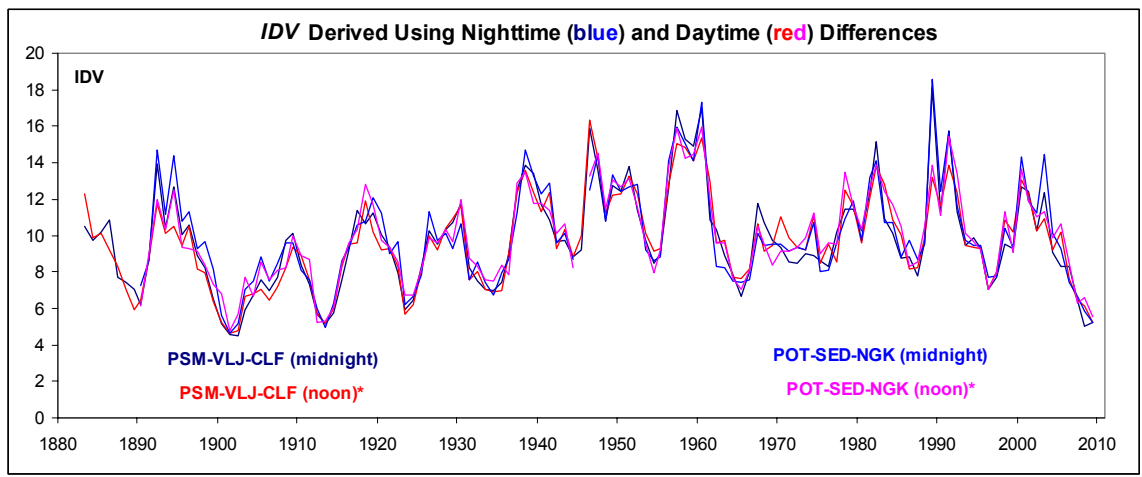

Figure 6. Raw *IDV* derived using night-time hourly data (blue) and daytime hourly data (red) for the French stations PSM-VLJ-CLF and the German stations POT-SED-NGK. The daytime values are ~30% larger than the night-time value because of day-to-day [non-solar] variations of the regular solar variations, $S_R$, and have been scaled to the same average as the night-time values for easier comparison.

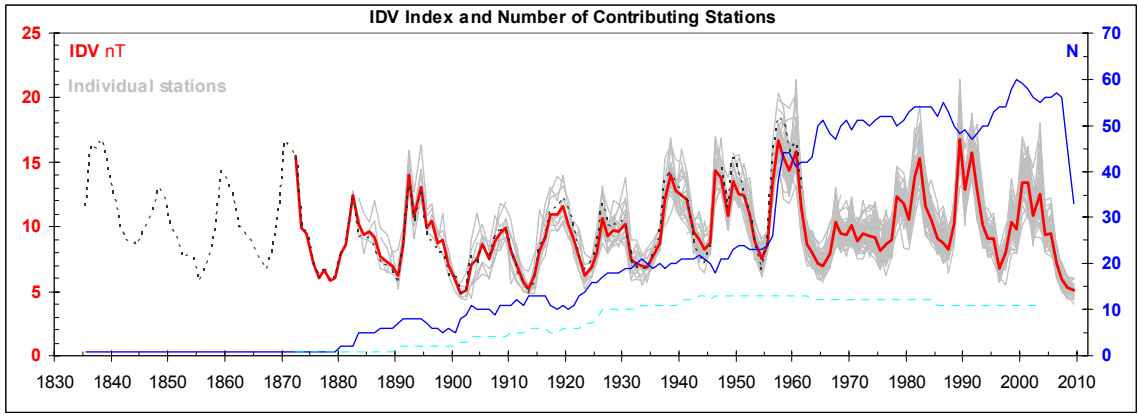

Figure 7. Yearly *IDV*-indices derived for individual stations (as given in Table 1) shown as grey curves. The red curve is a composite index calculated as the mean of the median and average values of the individual station values. This procedure may be justified by the very small difference between medians and averages (0.16 nT on average, see Figure 8). The number, *N*, of contributing stations is shown by the thin blue curve and the corresponding number for IDV05 as a dashed light blue line. The *u*-measure is considered a single station. A few station values differing more than five standard deviations from the average for a given year were omitted in calculating the average for that year. The dashed line shows *IDV* derived from the *u*-measure.

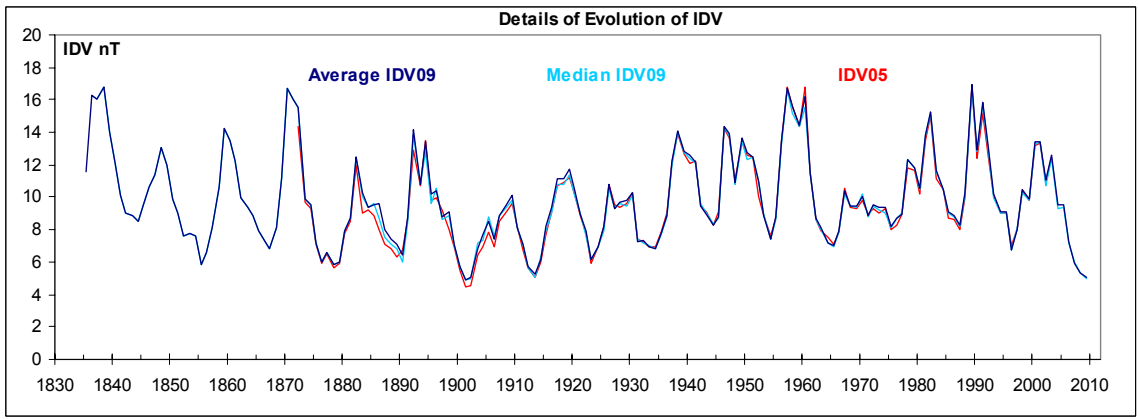

Figure 8. Average yearly values of IDV09 (dark blue curve) compared with median yearly values (light blue curve) and compared with published IDV05 (red curve).

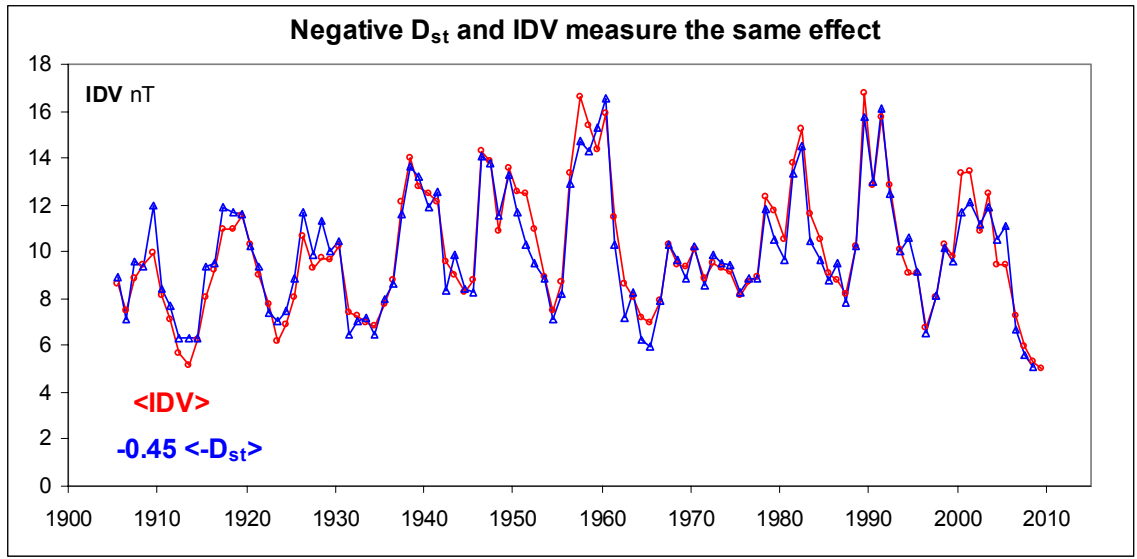

Figure 9. Yearly average values of *IDV* and of $D_{st}$ when it was less than zero (based on $D_{st}$ from Kyoto WDC and on $D_{st}$ from *Love* [2006] scaled to Kyoto levels). The 'spike' in 1909 is due to the extremely strong storm on 25 September 1909 causing loss of data at all but one station (API), giving that one data point undue influence. To guard against the influence of such sporadic extreme values, the daily values of *IDV* were capped at 75 nT.

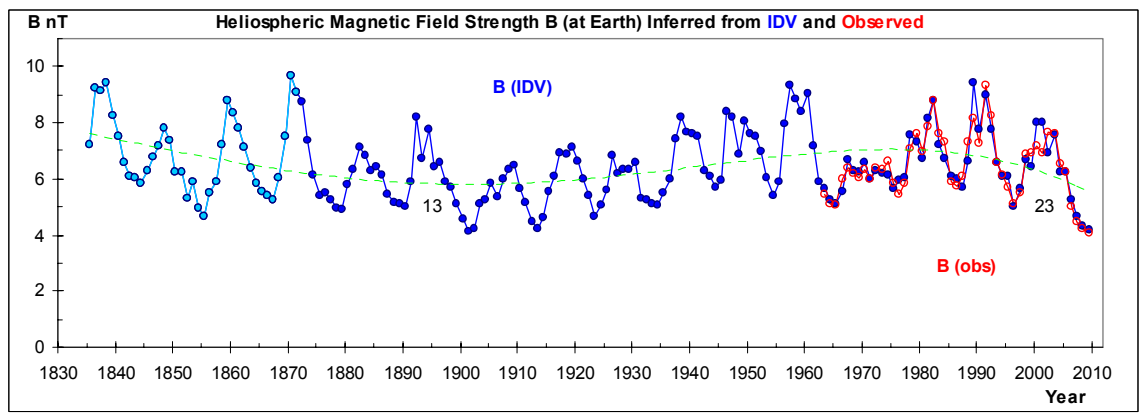

Figure 10. Yearly average values of the HMF *B* inferred from the *IDV*-index (dark blue curve) and from the early *u*-measure (light blue curve) compared with in situ measurements (red curve). There is a hint of the ~100 year Gleissberg cycle.

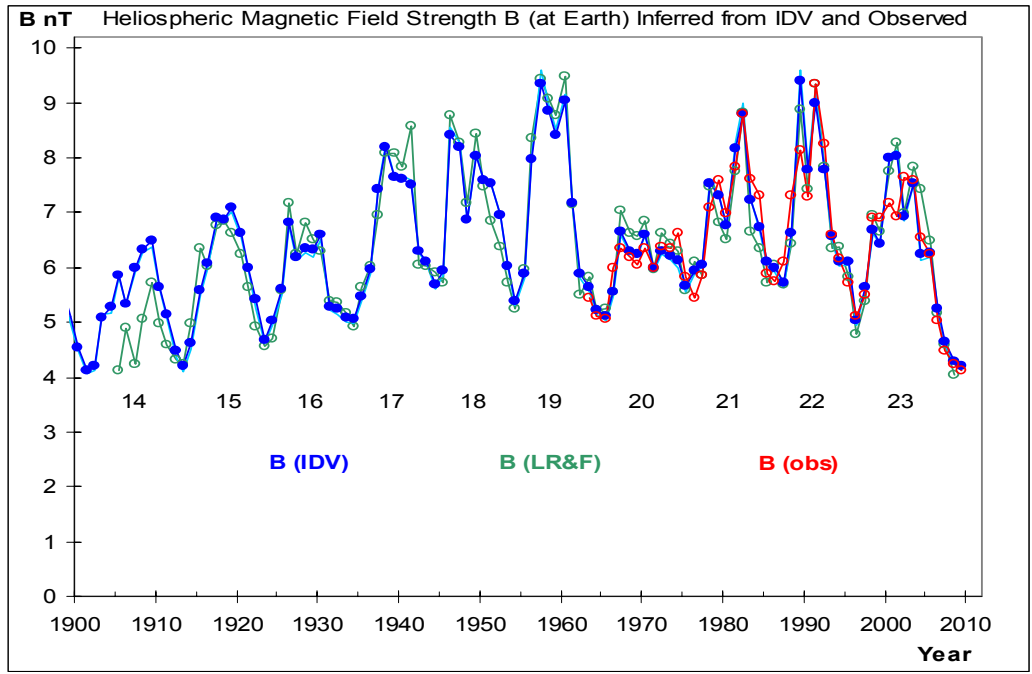

Figure 11. Comparison of HMF *B* determined from *IDV* [light blue curve using eq.(3) and dark blue curve using eq.(2)], by *Lockwood et al.* [2009, green curve], and observed by spacecraft [red curve].

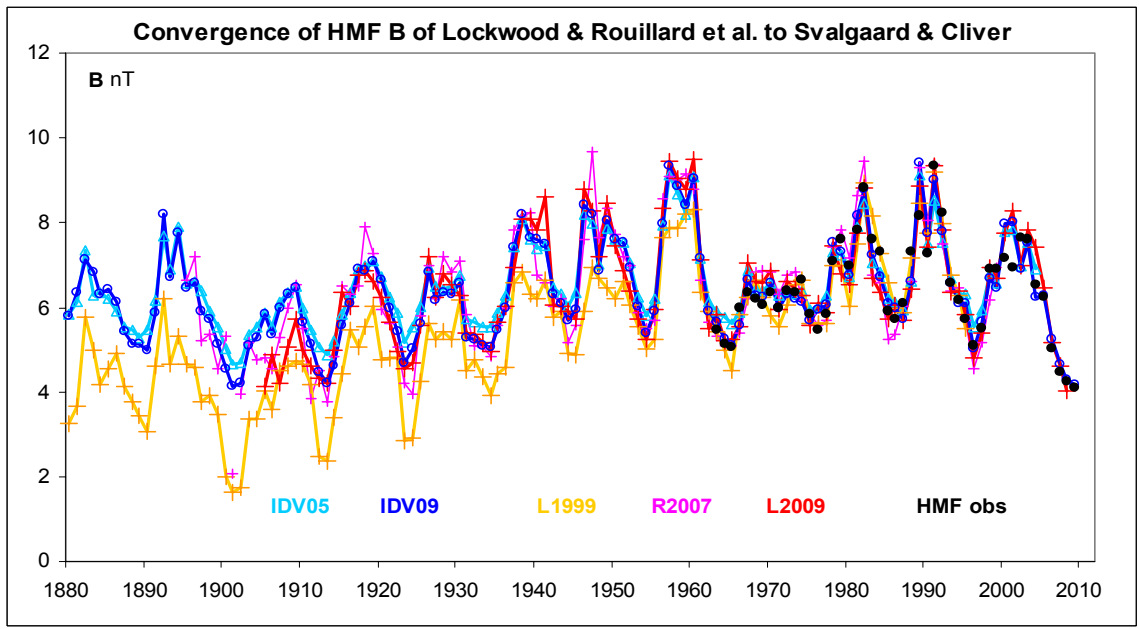

474  Figure 12. Comparison between HMF *B* derived by *Svalgaard and Cliver* [2005] (light
475  blue curve and open circles), this paper (dark blue curve and open circles) and HMF *B*
476  derived by *Lockwood et al.* [1999] (orange curve and plus-symbols), *Rouillard et al.*
477  [2007; their point for 1901 was in error, A. Rouillard, Personal comm. 2007] (pink curve
478  and plus symbols), and *Lockwood et al.* [2009] (red curve and plus-symbols), matched to
479  *in situ* observations of *B* (black dots).

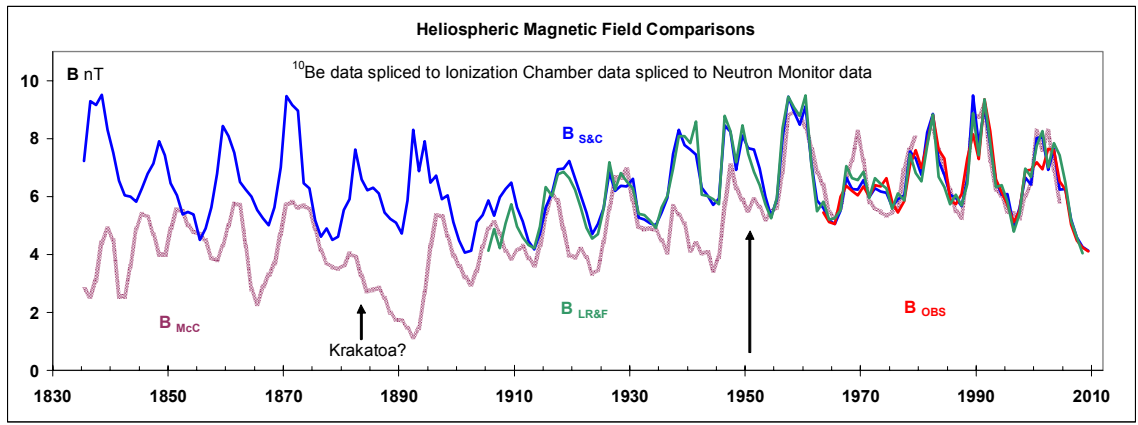

Figure 13. Yearly averages of near-Earth HMF *B* inferred by *Svalgaard and Cliver* [this paper] (blue curve $B_{S\&C}$), by *Lockwood et al.* [2009] (green curve $B_{LR\&F}$), observed by spacecraft (red curve $B_{OBS}$) compared to *B* inferred by *McCracken* [2007] (purple curve $B_{McC}$). The large arrow marks the beginning of the neutron monitor-based part of the record. One might speculate that the extremely low values during 1883-1893 are caused by the explosion of Krakatoa ejecting sulfur-rich aerosols into the stratosphere influencing the deposition of $^{10}$Be.

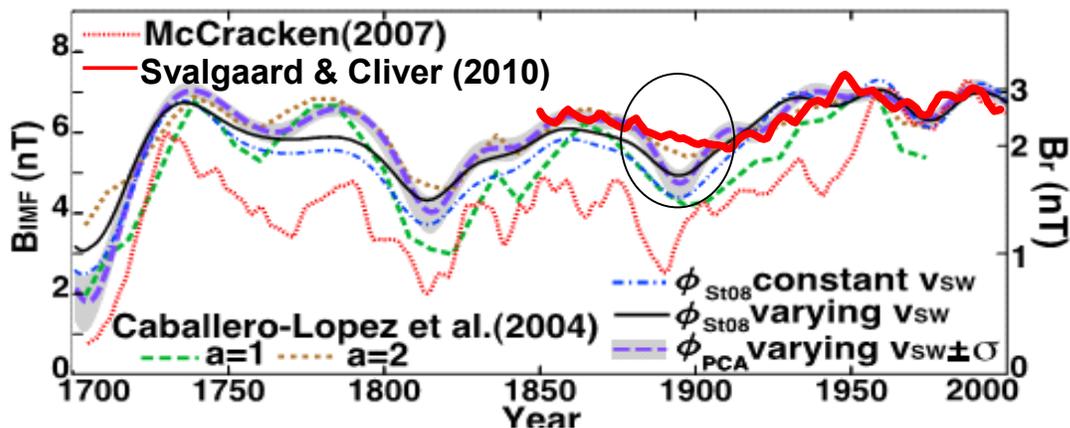

Figure 14. Comparison of our reconstruction of HMF *B* (red curve, 25 year running means) with other reconstructions as reported by *Steinhilber et al.* [2010] in their Figure 7 [adapted and reproduced here], *e.g.* with their 25-year running mean of their PCA-based reconstruction [purple dashed line] of *B*. The oval outlines an area of disagreement for which sufficient geomagnetic data exists that may be used to resolve the discrepancy.

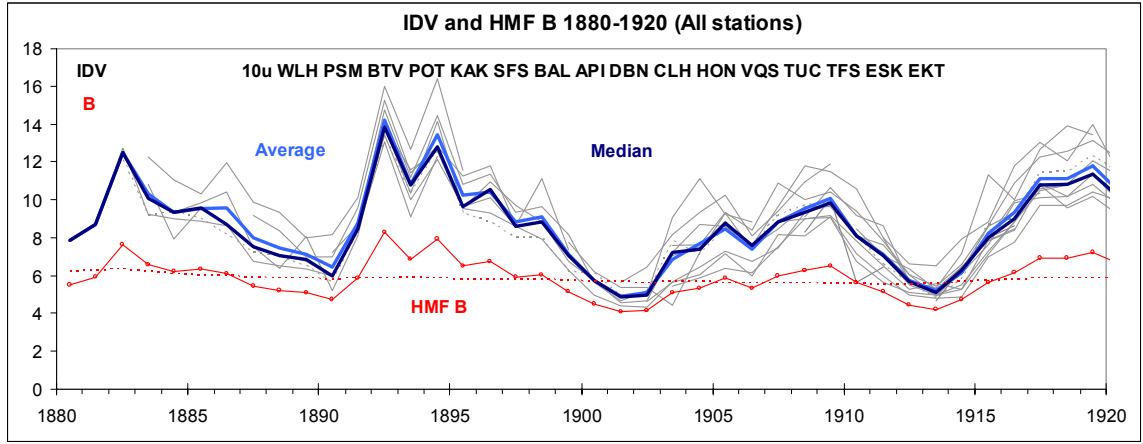

Figure 15. *IDV* [blue curves] and inferred HMF *B* [red curve; dashed line: 25-year running mean] 1880-1920 for all stations [as noted by their IAGA designations – 10*u* shown as a dashed gray line] where good geomagnetic data are available so far.